\documentclass[10pt,twocolumn,floatfix,prb,superscriptaddress]{revtex4-2}

\usepackage[dvipsnames]{xcolor}
\usepackage{amsmath,amssymb,amsthm,mathrsfs,amsfonts,dsfont,amstext} 
\usepackage{textcomp,pbox}
\usepackage[export]{adjustbox}
\usepackage{bm}
\usepackage{dcolumn,booktabs,url}
\usepackage[scaled]{helvet}
\usepackage{sansmath,gensymb}
\usepackage{tikz,graphicx,transparent}
\usepackage{multirow}
\usepackage[separate-uncertainty = true]{siunitx}
\usepackage{comment}
\usepackage{physics}
\usepackage{pdfpages}
\usepackage{array}
\usepackage{lineno}
\usepackage[normalem]{ulem}

\makeatletter
\AtBeginDocument{\let\LS@rot\@undefined}
\makeatother

\newcolumntype{C}[1]{>{\centering\let\newline\\\arraybackslash\hspace{0pt}}m{#1}}

\usepackage[colorlinks=true]{hyperref}
\usepackage{graphicx}

\hypersetup{
  colorlinks = true,
  citecolor = red,
   linkcolor  = blue,
   urlcolor   = red   
}

\graphicspath{{./figs/}}

\begin{document}
\newcommand{\TitleName}{
Dynamical photon-photon interaction mediated by a quantum emitter
}
\title{\TitleName}

\newcommand{\AffCPH}{Center for Hybrid Quantum Networks (Hy-Q), Niels Bohr Institute, University~of~Copenhagen, DK-2100 Copenhagen~{\O}, Denmark}
\newcommand{\AffBasel}{Department of Physics, University of Basel, Klingelbergstra\ss e 82, CH-4056 Basel, Switzerland}
\newcommand{\AffBochum}{Lehrstuhl f\"ur Angewandte Festk\"orperphysik, Ruhr-Universit\"at Bochum, Universit\"atsstra\ss e 150, 44801 Bochum, Germany}
\newcommand{\AffMAdrid}{Instituto de F\'{i}sica Fundamental IFF-CSIC, Calle Serrano 113b, Madrid 28006, Spain}

\author{Hanna Le Jeannic}
\altaffiliation[Present address: ]{Laboratoire Photonique Numérique et Nanoscience, Université de Bordeaux, Institut d’Optique, CNRS, UMR 5298, 33400 Talence, France}
\email{hanna.le-jeannic@cnrs.fr}
\affiliation{\AffCPH{}}

\author{Alexey Tiranov}
\affiliation{\AffCPH{}}

\author{Jacques Carolan}
\altaffiliation[Present address: ]{Wolfson Institute for Biomedical Research, University College London, Gower Street, London WC1E 6BT, United Kingdom}
\affiliation{\AffCPH{}}

\author{Tom\'{a}s Ramos}
\affiliation{\AffMAdrid{}}

\author{Ying Wang}
\affiliation{\AffCPH{}}

\author{Martin H. Appel}
\affiliation{\AffCPH{}}

\author{Sven Scholz}
\affiliation{\AffBochum{}}

\author{Andreas D. Wieck}
\affiliation{\AffBochum{}}

\author{Arne Ludwig}
\affiliation{\AffBochum{}}

\author{Nir Rotenberg}
\affiliation{\AffCPH{}}

\author{Leonardo Midolo}
\affiliation{\AffCPH{}}

\author{Juan Jos\'{e} Garc\'{i}a-Ripoll}
\affiliation{\AffMAdrid{}}

\author{Anders S. S\o rensen} 
\affiliation{\AffCPH{}}

\author{Peter Lodahl}
\email{lodahl@nbi.ku.dk}
\affiliation{\AffCPH{}}

\date{\today}

\begin{abstract}
Single photons constitute a main platform in quantum science and technology: they carry quantum information over extended distances in the future quantum internet\cite{Kimble2008} and can be manipulated in advanced photonic circuits enabling scalable photonic quantum computing \cite{Wang2020,Uppu2021}. A key challenge in quantum photonics is how to generate advanced entangled resource states and efficient light-matter interfaces offer a path forward\cite{Lindner2009,Chang2014}. Here we utilize the efficient and coherent coupling of a single quantum emitter to a nanophotonic waveguide for realizing quantum nonlinear interaction between single-photon wavepackets. This inherently multimode quantum system constitutes a new research frontier in quantum optics \cite{Chang_RMP_2018}. We demonstrate the control of a photon using a second photon mediated by the quantum emitter. 
The dynamical response of the two-photon interaction is experimentally unravelled and reveals quantum correlations controlled by the pulse duration. 
This work will open new avenues for tailoring complex photonic quantum resource states.

\end{abstract}

\maketitle 

The interaction of a single quantum of light and a single quantum emitter has been a long-standing endeavour in quantum optics \cite{Haroche2006}. The envisioned quantum-information applications range from photon sources \cite{Kimble1977,Kuhn2002} to photonic quantum gates \cite{Duan_Kimble_2004,Reiserer_2014}. The paradigmatic setting captured by the Jaynes-Cummings model \cite{Haroche2006,Jaynes1963} describes a single confined optical mode interacting with a single quantum emitter. Photonic cavities enable fast and controllable single-photon switching\cite{Englund2012,Sun2018,Volz2014}, and near-deterministic and coherent light-matter coupling has been reported \cite{Najer2019,Pscherer2021}.
Recently, waveguide quantum electrodynamics (WQED) has emerged where the quantum emitter is coupled to a travelling mode of light \cite{Lund-Hansen2008,Chang2007,Vetsch2010, Tey2008,Lang_2011,Deppe2008,Goban2014,Peyronel2012}. This inherently open quantum system constitutes a new paradigm in quantum optics \cite{Lodahl2015,Chang_RMP_2018} enabling chiral quantum optics \cite{Lodahl2017}, topological photonics \cite{Barik2018}, and fundamentally new bounds on quantum optics devices \cite{Asenjo2017}. WQED systems constitute an attractive photon-emitter interface since they realize a wide optical bandwidth with near-deterministic coupling efficiency \cite{Arcari2014}, which is advantageous when studying quantum pulses interacting with the emitter.

At its most fundamental level, WQED features a single quantum emitter coupled to a continuum of optical modes forming a quantum pulse \cite{Kiilerich2019}. The quantum complexity of this nonlinear system spanning a multi-dimensional Hilbert space is remarkable \cite{Fan2010}, and complex physical phenomena have been proposed and analyzed theoretically, including photonic bound states \cite{Shen2007a,Mahmoodian2020}, the generation of Schr\"{o}dinger cat states \cite{Kiilerich2019}, and stimulated emission in the most fundamental setting of one photon stimulating one excited emitter \cite{Rephaeli2012}. Previous work focused on the monochromatic case where photon-photon interaction was realized \cite{Javadi2015,LeJeannic2020}. Here we experimentally demonstrate quantum nonlinear interaction between few-photon pulses mediated by the interaction with a single quantum emitter in a waveguide. 

\section{Main}

\begin{figure*}[h!]
	\includegraphics[width=1\linewidth]{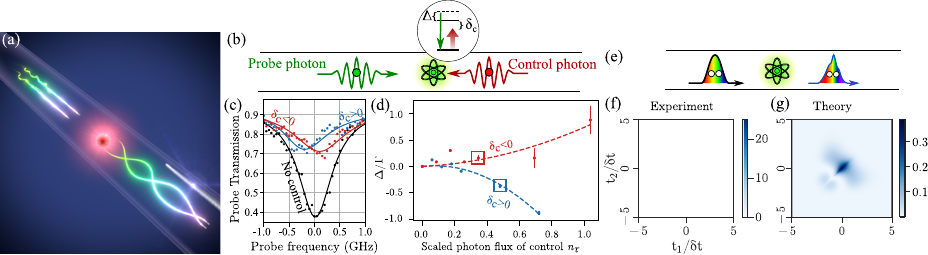}
	\caption{Observation of dynamical photon-photon interaction (color online).
(a) Conceptual illustration of two-photon pulsed nonlinear interaction mediated by a quantum emitter in a nanophotonic waveguide inducing strong quantum correlations between the photon wavepackets. (b) Quantum control experiment where the interaction (transmission/reflection) of a single probe photon (green) with the quantum emitter is controlled by another photon (red). The control photon effectively shifts the emitter resonance by an amount $\Delta$ that can be controlled by the detuning of the control photon from the bare resonance $\delta_c$ and the control photon flux. (c) Measured transmission of the probe beam in the absence of a control photon (black curve) and with a control signal of $\delta_c = \omega_c - \omega_0 =-0.3\Gamma$, $n_{\tau}=0.24$ (red curve) and $\delta_c = 0.3\Gamma,$ $n_{\tau}=0.7$ (blue curve). 
(d) Measurement of the resonance shift $\Delta/\Gamma$ versus the scaled number of control photons for $\delta_c = 0.3\Gamma$ (blue data) and $\delta_c = -0.3\Gamma$ (red data). The red and blue boxes indicate the data displayed in (c). Less than one photon suffices <to shift the resonance by a full linewidth. The input intensity corresponds to $\approx 2$ times the saturation level.
(e) Illustration of temporal quantum correlations induced by the interaction of two single photons via the quantum emitter. (f) Experimentally recorded second-order correlation function in the transmission geometry for a Gaussian pulse of duration (standard deviation) of $\delta t =340$~ps after interaction with the quantum dot. (g) Calculated second-order transmission correlation function for a two-photon pulse of duration $\delta t =340$~ps and the ideal case of a quantum emitter deterministically coupled to the waveguide without any imperfections. 
}
	\label{figure1}
\end{figure*}

Figure \ref{figure1}(a) shows the conceptual setting of the experiment: two quantum pulses propagate in the waveguide and interact with a single quantum emitter. If the photon-emitter coupling cooperativity is high \cite{Lodahl2015}, even a single photon interacts efficiently with the emitter and can ultimately saturate it. This results in the reflection of a single photon \cite{Chang2007}. Consequently, two simultaneous photons are strongly transformed by the interaction with the emitter, effectively leading to photon-photon nonlinear interaction. Two different experimental settings are realized: i) one photon in the waveguide can control the transmission of another, see Fig.~\ref{figure1}(b), in a single-photon version of pump-probe spectroscopy experiments traditionally requiring high photon fluxes \cite{Wu1977}
 ii) two-photon pulsed interaction where the strong interaction with the emitter induces complex temporal quantum correlations, see Fig.~\ref{figure1}(e). Realizing such fundamental quantum nonlinear processes requires a quantum coherent and highly-efficient light-matter interface, which is obtained using a semiconductor quantum dot in a photonic-crystal waveguide. Quantum nonlinear optics has been previously studied on different experimental platforms, including solid-state defect centers \cite{Sipahigil2016}, atoms \cite{Vetsch2010,Goban2014}, molecules \cite{Maser2016}, quantum dots \cite{Javadi2015}, and micro-wave resonators \cite{Mirhosseini2019}, but experiments were mainly limited to monochromatic excitation, i.e. the rich dynamics of quantum pulses has remained largely unexplored.

First consider the two-color photon-photon control experiment; a primer in quantum nonlinear optics\cite{Maser2016}.
Figure \ref{figure1}(c)+(d) displays the experimental data showing how a control beam of frequency $\omega_c$ launched through the waveguide effectively shifts the quantum dot by an amount $\Delta$ depending on the photon flux and $\omega_c$, which is the AC Stark-effect\cite{Maser2016}. The proof-of-concept experiment exploits a monochromatic weak coherent laser, and the single-photon sensitivity is realized by observing that on average less than a single photon (within the quantum-dot lifetime) suffices to shift the resonance by a significant fraction of the radiative linewidth $\Gamma$. We find that a scaled photon flux of $n_{\tau}=0.97 \pm 0.27$ (average number of photons within the emitter lifetime) detunes the quantum dot by a full linewidth, see Methods for the flux calibration analysis. Consequently, a control photon modulates the probe photon that is either preferentially reflected ($\Delta=0$) or transmitted ($|\Delta|> \Gamma$). $\Delta$ and the switching contrast (see Supplemental Material) change with the photon flux of the control beam and with its detuning $\delta_c = \omega_{c} -\omega_{0}$ from the bare quantum dot resonance $\omega_{0}$. These two parameters therefore constitute "control knobs" of the photon-photon interaction, see Fig.~\ref{figure1}(d). 
 We note that this quantum switch operates with an intrinsic timescale determined by the lifetime of the quantum dot (sub-nanoseconds) and may find practical applications in quantum photonics or deep learning using nanophotonics where fast optical switching is a key requirement \cite{Shen2017,Uppu2021}.

To access the temporal dynamics of the non-linearity we study the two-photon nonlinear response by recording the second-order intensity correlation function $C_{tt}^{(2)}(t_1,t_2)$ for weak coherent Gaussian pulses (See Supplemental Material and Ref. \cite{Ramos2017} for more details). Here $t_1, t_2$ are the photon detection times and subscript $t$ indicates that both photons are detected in the transmission channel, see Ref. \cite{LeJeannic2020} for a description of the experimental approach. 
Figure \ref{figure1}(f) shows a representative experimental data set. A complex temporal quantum correlation structure is observed, as witnessed by the ``bird-like'' image reflecting that the incoming photon wavepacket is reshaped through the nonlinear interaction by an amount depending on the photon number. The detailed one- and two-photon dynamical response is mapped out below. For comparison, Fig.~\ref{figure1}(g) shows the calculated second-order intensity correlation function in the ideal case of a fully deterministically and coherently coupled quantum dot, i.e., the ideal ``1D quantum emitter'' with no residual radiative loss or decoherence. The calculation of the two-photon response was obtained following an approach as outlined in Ref \cite{Heuck2020}. 
Remarkably, the resemblance of the experimental data to this ideal case testifies the high performance of the system and the ability to map out the two-photon response. In the following we will unravel the underlying dynamics of the photon-emitter interaction processes.

\begin{figure*}[ht]
	\includegraphics[width=1.0\linewidth]{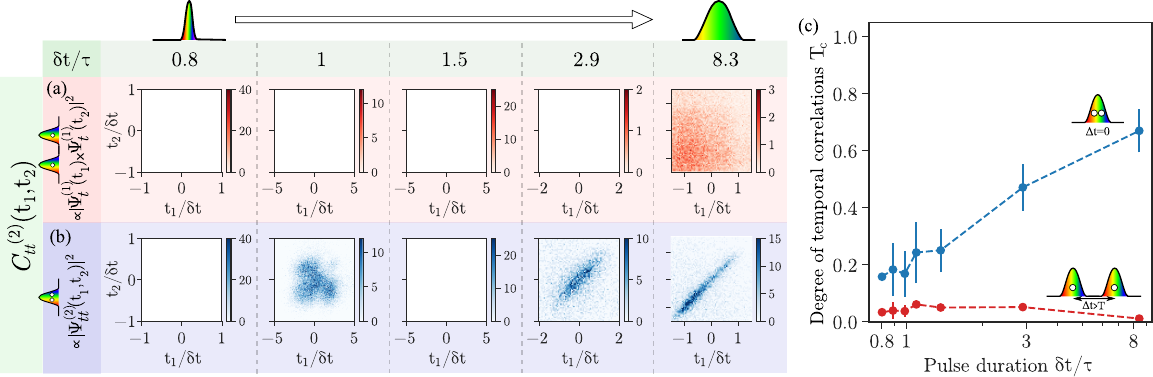}
	\caption{Temporal quantum correlations due to photon-photon dynamical interaction in the transmission channel (color online).
	 Measured time-resolved second-order correlation function for various pulse duration $\delta t$ relative to the quantum dot lifetime $\tau$ and for (a) two single photons from subsequently scattered pulses and (b) two photons contained in the same pulse. (c) Extracted degree of temporal correlation $T_c$ versus pulse duration for the two data sets (a) [red] and (b) [blue]. 
		}
	\label{figure2}
\end{figure*}

The two-photon dynamics is explored in Fig.~\ref{figure2} by recording the two-time correlation function in transmission for different durations of the incoming pulse, $\delta t$, relative to the emitter lifetime, $\tau\approx 229$~ps (see Methods).
Two interaction processes are compared, depending on the temporal separation between pulses: i) independent scattering of temporally separated single-photon pulses from the quantum dot (Fig.~\ref{figure2} (a)) and ii) two-photon scattering of photons originating from the same pulse (Fig.~\ref{figure2} (b)). 
Experimentally both cases can be extracted from a single series of pulsed two-photon correlation functions by analyzing data from i) subsequent pulses ($t_2 \approx t_1+\Delta t$, where $\Delta t=30$~ns $\gg \tau$ is the delay between excitation pulses) or ii) same pulses ($t_2 \approx t_1 $). The nonlinear interaction induces temporal quantum correlations on a time scale determined by the pulse duration $\delta t$ and the lifetime $\tau.$ 

Case i) of independent single-photon scattering serves as a reference measurement essentially corresponding to an uncorrelated case. 
The two input pulses are separated by more than the lifetime, i.e. the emitter does not mediate any photon-photon interaction. The correlation measurements probe single-photon (denoted by superscript 1 on wavefunction $\Psi$) components of the scattered wavefunction, i.e. $C_{tt}^{(2)}(t_1,t_2)\propto |\Psi^{(1)}_t(t_1)|^2 |\Psi^{(1)}_t(t_2)|^2$, see Fig.~\ref{figure2}(a). The observed correlation plots can therefore be interpreted from single-photon dynamics. A short input pulse, $\delta t/\tau\lesssim 1$, is spectrally wide and has therefore a small overlap with the quantum dot bandwidth meaning that the pulse is preferentially transmitted with little effect from the emitter. Increasing the pulse duration, $\delta t/\tau\gtrsim 1$, increases the interaction with the quantum dot and thereby the probability to reflect a single photon from the incoming pulse. This reduces the probability of photon transmission (observed as a low probability amplitude around $t_1\sim t_2\sim 0$), which results in the visible "cross-like" destructive interference, and the overall transmission probability reduces as the pulse duration grows further. 

Case ii) reveals the dynamics of two-photon (superscript 2 on wavefunction) scattering processes, i.e. $C_{tt}^{(2)}(t_1,t_2)\propto |\Psi_{tt}^{(2)}(t_1,t_2)|^2$. The quantum dot mediates strong photon-photon correlations tailored by the duration of the incoming pulse. For $\delta t/\tau\lesssim 1$, the pulse is spectrally wide, and only weak interaction is observed similar to case i), see data in Fig.~\ref{figure2}(b).
For longer pulses, $\delta t/\tau\gtrsim 1$, the interaction increases and we observe strong temporal correlation, i.e. the detection of one photon increases the probability of detecting another. 
This is observed in Fig.~\ref{figure2}(b) as the clustering of data points around the axis $t_1=t_2$ for long pulses. 
The observed photon bunching in the transmission channel stems from the fact that the quantum dot can only scatter one photon at a time, and was observed previously only in continuous-wave experiments \cite{Javadi2015,Liang2018}. The present experiment reveals the dynamics of this nonlinear photon-sorting process.

The temporal correlations can be quantified by performing a Schmidt decomposition of the experimental data $C^{(2)}_{tt}(t_1,t_2)$ \cite{zielnicki2018joint,Law2000} (see Methods for details).
From the Schmidt coefficients $\lambda_i$ we extract the degree of temporal correlation $T_c=1-\sum_i \lambda_i^4$ versus pulse duration $\delta t/\tau$, see Fig.\ref{figure2}(c). Case i) of independent scattering does not introduce any significant correlations, $T_c \simeq 0$, which is the case for a separable quantum state. A fundamentally different behavior is observed for the two-photon scattering case of ii) where $T_c$ is found to grow with pulse duration. This behavior, sensitive to the coherence of the emission and to the coupling efficiency, is a manifestation of the observed correlated photon-pair emission (see Fig.\ref{figure2}(b)) resembling nonlinear parametric down-conversion or four-wave mixing sources\cite{Kuzucu2008}. In the present implementation, a single quantum dot deterministically coupled to a waveguide acts as the photon-pair source. 

\begin{figure*}[ht]
	\includegraphics[width=1.0\linewidth]{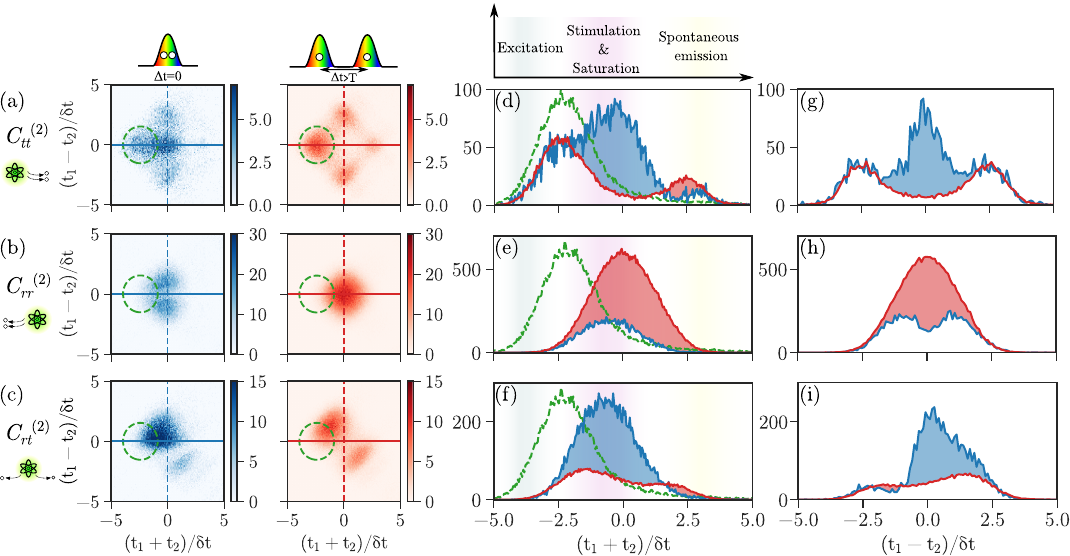}
	\caption{Unravelling the physical processes behind the quantum dynamics (color online). Experimental measurements of the two-photon correlation function for $\delta t/\tau = 1.5$ and photons detected in different spatial modes i.e. both photons being transmitted (a), reflected (b), or one photon reflected and the other transmitted (c) by the quantum dot. The two different cases correspond to two photons in the same pulse (blue data) or one photon in each subsequent pulse (red data). The green dashed line marks the position of the incident pulse used to excite the quantum dot.
	(d)-(f) Line cuts at $t_1=t_2$, indicated by the full line in the correlation data in (a)-(c), as a function of $t_1 + t_2$ for the three cases. The main physical processes responsible for the dynamics in the various temporal domains are noted.(g)-(i) Line cuts at $t_1=-t_2$, indicated by the dashed line in the correlation data in (a)-(c) as a function of $t_1 - t_2$ for the three cases. To increase the signal-to-noise ratio of the line cuts, the coincidence counts have been integrated over 10 bins (corresponding to 200 ps).
		}
	\label{figure3}
\end{figure*}

 The WQED photon-photon nonlinear interaction has unique features arising from an intricate interplay between the drive pulse and the field scattered by the quantum dot, resulting in quantum interference spread over diverse spatial degrees of freedom. We examine different propagation directions of the output field through the simultaneous recording of $C^{(2)}_{\mu\mu'}(t_1,t_2)$ for reflection or transmission channels $\mu,\mu'=t,r$, and comparing both the one- and two-photon cases (see Fig.~\ref{figure3}(a)-(c)). We apply a weak drive pulse ($\ll 1$ photons on average), to avoid three or higher-order photon processes. In the forward propagating direction (transmission channel) quantum interference is present, while in the backward direction (reflection channel), solely the scattering response of the quantum dot is observed. Furthermore, the cross-correlation between reflection/transmission channels is also studied. Line-cuts through the two-dimensional correlation plots are presented both versus the sum of the detection times (Fig.~\ref{figure3}(d)-(f)) and versus the delay (Fig.~\ref{figure3}(g)-(i)) comparing both the one-photon and two-photon responses. These data sets are instructive for the physical interpretation of the quantum dynamics.
 Three different regimes are defined corresponding to: 1) excitation, 2) saturation and stimulated emission, and 3) spontaneous emission of the emitter. 
 
 In regime 1), the polarization of the emitter builds up due to the rise of the excitation pulse. Here the one- and two-photon dynamics are similar since the probability of absorption remains small. The build-up of the excitation probability is directly revealed in the reflection data (Fig.~\ref{figure3} (e)), since no interference with the incoming pulse occurs in this case. 
 
 As the excitation probability becomes sizable, we enter regime 2) of stimulated emission and saturation, where stark differences between one- and two-photon dynamics are observed. The reflection is strongly suppressed in the two-photon case (Fig.~\ref{figure3} (e)). This is a direct consequence of the emitter only reflecting one photon at a time, leading to the observable dip in the time delay data in Fig.~\ref{figure3} (h). The single-photon response is dominated by a strong reflection, a testimony of the efficient coupling of the emitter to the waveguide leading to a large optical extinction, which is confirmed by the suppression of transmission-transmission and transmission-reflection events in Fig.~\ref{figure3} (d) and (f), respectively. In contrast, a pronounced enhancement is found for the two-photon dynamics since a single photon suffices for saturating the emitter enabling the transmission of a second photon. The time delay data in Fig.~\ref{figure3} (g) and (i) allow to further discern the dynamics of this process. The strong asymmetry in the transmission-reflection data (Fig.~\ref{figure3} (i)) reveals the temporal ordering of the process, where a photon is first absorbed, then a second photon is transmitted and finally the first photon is re-emitted. In the transmission-transmission channel, the two detected photons had propagated in the same direction enabling stimulated emission. We observe a pronounced preference for two-photon transmission compared to the single-photon case, see Fig.~\ref{figure3} (d). We further monitor the delay between the transmitted photons, and find an increased emission rate in the forward (transmission) direction by comparing the time delay data in Fig.~\ref{figure3} (g) to the transmission-reflection data in Fig.~\ref{figure3} (i). These observations are signatures of stimulated emission of a saturable emitter occurring here in the most fundamental setting of just two quanta of light and mediated by a single quantum emitter. Indeed, with the efficient and coherent photon-emitter coupling in the photonic-crystal waveguide, even a single photon pulse suffices for stimulating emission.
 
 Finally, after the excitation pulse has passed, the system enters into regime 3) where the remaining population of the emitter decays by spontaneous emission. We observe that generally the two-photon response is suppressed relative to the one-photon response reflecting the fact that the single emitter only stored one excitation. 
The duration and effect of those three regimes depend on the pulse duration compared to the emitter response time, and similar data for different pulse lengths can be found in the Supplemental Materials

Using a quantum dot deterministically coupled to a nanophotonic waveguide, we have reported two fundamental demonstrations of quantum nonlinear optics: a single-photon pump-probe experiment where one photon controls another and a quantum-pulse experiment where photon-emitter dynamic scattering was discerned into its most fundamental constituents. The current focus was to unravel the underlying physical processes behind quantum nonlinear interaction with quantum pulses, however applications are foreseen. For instance, photon sorters have been proposed as a basis for a deterministic Bell analyzer for photons \cite{Witthaut2012}, which is a key enabling component in photonic quantum information processing. Another interesting direction is to exploit and tailor the nonlinear interaction to synthesize specific photonic quantum states \cite{Kiilerich2019}, possibly boosted in a quantum optics neural network \cite{Steinbrecher2019}. Hybrid discrete-continuous variable architectures for photonic quantum computing appear another promising future research direction, since the nonlinear response of the emitter could provide a non-Gaussian photonic operation, which is currently the "missing link" in continuous-variable quantum information processing.

\section{Data Availability}
The data that support the findings of this work are included in the Supplementary Information. This includes the complete dataset of time correlation measurements for different pulse lengths, in all of the three propagation directions (Supplementary Figures 1 and 2). Further data are available from the corresponding author upon reasonable request.

\section{Acknowledgments}
We thank Klaus M\o lmer and Ravitej Uppu for valuable discussions.
 We acknowledge funding from the Danish National Research Foundation (Center of Excellence “Hy-Q,” grant number DNRF139). This project has received funding from the European Union’s Horizon 2020 research and innovation programme under grant agreement No. 824140 (TOCHA, H2020-FETPROACT-01-2018).

\bibliography{biblio}

\newpage
\section{Methods}
\subsection{Photon-emitter interface}
The considered quantum emitter is a neutral excitonic state of a self-assembled InGaAs quantum dot (QD). The emitter is embedded in a GaAs suspended photonic-crystal waveguide and includes doped layers to form a p-i-n diode heterostructure, which enables electrical contacting allowing charge stabilization of the environment and tuning of the resonance through the DC Stark effect. Details about the sample can be found in Ref \cite{Kirsanske2017}. The sample is cooled down to $4~$K to reduce phonon-induced dephasing.
The transition decay rate has been measured through p-shell excitation to be $4.364(5)~$ns$^{-1}$, corresponding to a measured lifetime of $\tau\simeq 229~$ps. For comparison, the linewidth of the transition is measured to be $755~$ MHz.

\subsection{Two-color photon control experiment}
\subsubsection{Experimental setup}
\begin{figure}[ht]
	\includegraphics[width=0.9\linewidth]{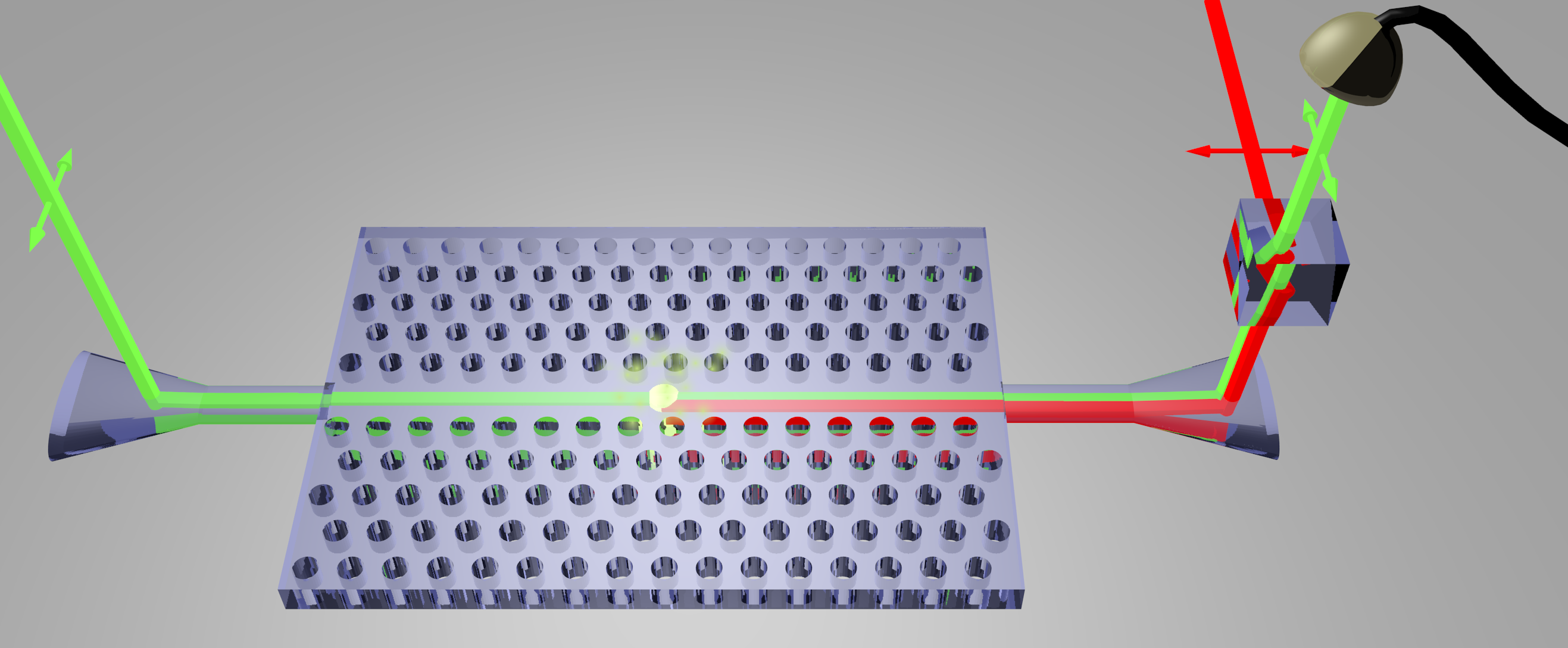}
	\caption{Experimental setup for the two-color photon control experiment. The probe (green) and control (red) CW weak lasers of orthogonal polarizations are sent through the waveguide by using opposite gratings. The transmission of the probe signal can be recorded with single photon detectors.}
	\label{Methods_setup_2colors}
\end{figure}

In the experiment realizing two-photon control, see data in Fig.~\ref{figure1}(c)+(d) of the main manuscript, two tunable continuous-wave (CW) lasers (linewidth $<10~$kHz) were applied for the probe and control s that excited the QD through the two gratings of the nanophotonic waveguide, see sketch in Fig.~\ref{Methods_setup_2colors}.
By using a combination of polarisation optimization and careful alignment, the transmission of the probe signal was recorded, with an extinction ratio between the laser excitation and the signal of $\approx 15$.

\subsubsection{Calibration of the control photon flux}

\begin{figure}[ht]
	\includegraphics[width=0.9\linewidth]{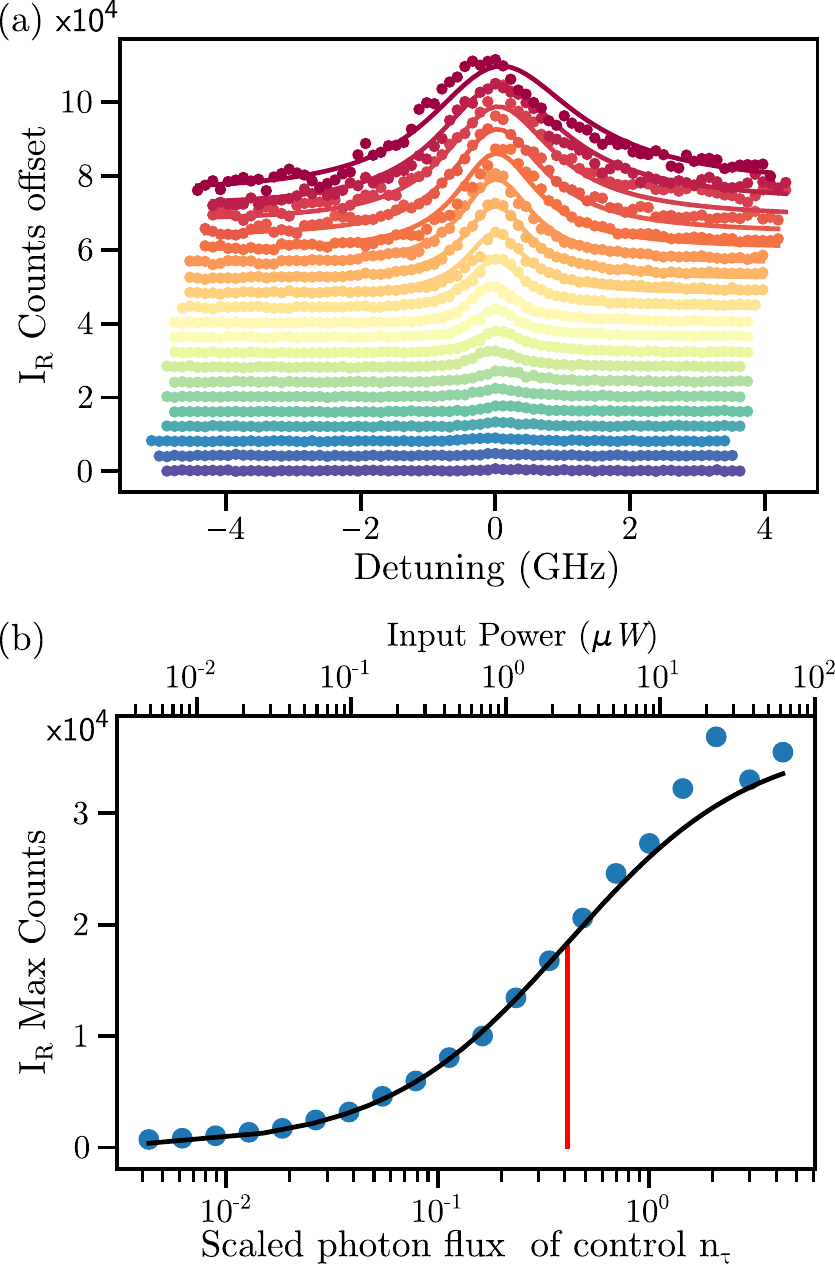}
	\caption{Calibration of photon flux. (a) Fluorescence measured in reflection (points) and corresponding fit to theory as a function of laser power (going from low (blue) to high (red) power) versus detuning. Fits are shown with solid lines and the fluorescence is offset as a function of laser power for clarity. (b) The measured intensity $I_R$ as a function of scaled photon flux for measured data (points) alongside the theoretical curve. Error bars due to Poissonian counts are smaller than the point size. The critical photon number $n_c\approx 0.42$ is marked, demonstrating that an efficient optical nonlinearity with single-photon sensitivity is achieved.}
	\label{figure_2color}
\end{figure}

To determine the number of photons required to switch the QD, we first calibrate the control laser power in the waveguide by recording a saturation measurement of the QD. The fluorescence intensity spectrum $I_R$ reflected by the QD is measured as a function the QD-laser detuning $\Delta$ and laser power $P$, see data in Fig.~\ref{figure_2color}(a).
The counts $I_R$ are corrected for background and the spectra are fitted using the formulas derived in Ref.~\cite{LeJeannic2020}.
For modelling the data, we used the following set of parameters: $\beta \approx 0.9$, dephasing rate $\Gamma_0 \approx 0.3$~ns$^{-1}$, and the calibrating parameter $\alpha \approx 0.3 $~ns$^{-2}/\mu W$ relates the Rabi frequency $\Omega$ to the laser power through $\Omega = \sqrt{\alpha P}$. The decay rate of the emitter was independently measured to be $\Gamma_{\text{tot}} = 4.364(5)$~ns$^{-1}$. 

The critical photon flux during one lifetime of the control beam is then calculated to be:
$n_c = (1 + 2\Gamma_0/\Gamma_{\text{tot}})^2 / 4 \beta^2$, which for our system was determined to be $n_c\approx0.42$\cite{Javadi2015,Thyrrestrup2018}.
We can finally calibrate the scaled photon flux of the control beam $n_{\tau}$ by using: $n_{\tau} = S n_c$, where the saturation parameter is then given by $S = 8 \Omega^2/(\Gamma_{\text{tot}}(2\Gamma_0 + \Gamma_{\text{tot}}))$
As a sanity check, we can compare the measured $I_R$ against the analytic form of the saturation curve at resonance: $I_R = a \beta^2 \Gamma_\text{tot}^2 /(8S)$, which is shown in Fig.~\ref{figure_2color}(b).

\subsubsection{Extracting the nonlinear resonance shift}
\begin{figure}[ht]
	\includegraphics[width=0.9\linewidth]{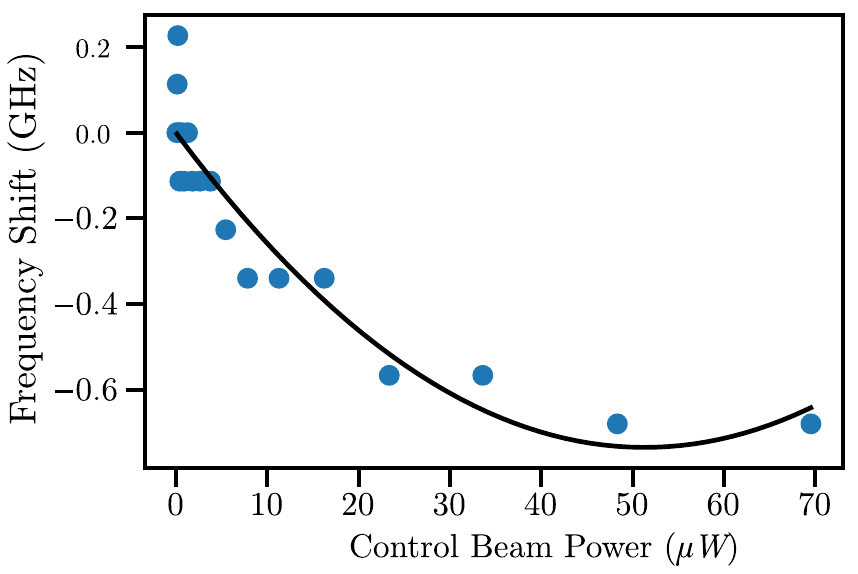}
	\caption{Frequency calibration measurements. The shift in the resonance frequency of the fluorescence of the QD as a function of control laser power (blue points), fitted to a second order polynomial (black line).}
	\label{figure_2color_freq_shift}
\end{figure}

 To calculate the nonlinear resonance shift, the control beam is detuned $\pm 200$~MHz relative to the QD, the probe beam is scanned across the QD and the transmission measured as a function of control laser power.
The control beam naturally induces a power-dependent frequency shift, always towards longer wavelengths in the QD due to thermal effects and carrier creation. To account for this, and to isolate the true multi-color nonlinear effect, we pre-characterize the power-dependent frequency shift of the control laser before each measurement.
 Fig.~\ref{figure_2color_freq_shift} shows an example of the measured QD frequency shift versus control laser power constituting a calibration curve. 
We then apply a power-dependent frequency correction to the control beam to maintain the QD-control beam detuning, effectively `tracking' the QD as a function of power.
The probe beam transmissions are then fitted by a Lorentzian function to estimate the central frequency, which is plotted as a function of photon flux in Fig.~\ref{figure1}(d). 
A second-order polynomial is fitted to the photon number versus normalised frequency shift. From this, we can determine the photon flux required to shift the QD by a full linewidth $\Delta / \Gamma_\text{tot} = -1$, when the control beam is detuned by $200$ MHz relative to the resonance. We find $0.97 \pm 0.27$ photons within the emitter lifetime shift the QD by a full linewidth. This corresponds to a saturation parameter of $S=n_\tau/n_c\approx 2.3$.

\subsection{Dynamics of two photons interacting with a quantum emitter}
\subsubsection{Experimental setup}
In the second experiment, the temporal photon-photon dynamics is probed. Here a CW laser (linewidth $<10~$kHz) is sent to a $20$~GHz electro-optical modulator (iXBlue NIR-MX800-LN-20) to generate tunable pulses with duration between 300~ps and 10~ns. Furthermore, 100~ps pulses are generated using another pulse generator (Alnair EPG-210 picosecond electrical pulse generator) and an external clock. The repetition rate of the experiment is set to $33~$MHz, enabling time delay between the pulses much longer than the emitter's response time. The laser central wavelength is tuned to the resonance of the exciton and is strongly attenuated to contain an average photon number below $0.1$ photons within the lifetime of the emitter.
Two-photon correlation measurements are performed in the different propagation directions of the light, following the same scheme as detailed in Ref \cite{LeJeannic2020} for CW-excitation. The coincidence events are detected with four superconducting nanowire single photon detectors (SNSPD) with timing jitters below $30$~ps in transmission, and below $150$~ps in reflection, and using a Swabian ultra time tagger. To avoid issues related to the accumulation of jitter over long time acquisition, the clock signal of the laser is also registered, and single photon time detection events are registered according to this clock signal. 

In a single measurement run, we are able to access both the correlation data originating from one-photon and two-photon interactions. This is done by recording the second-order intensity correlation function $G^{(2)}_{xy}(t_1,t_2)$ with two single-photon detectors in a pulsed experiment. By recording two-photon detection events where $t_1\approx t_2$ and $t_1\approx t_2 + \Delta t$, respectively, we post-select on the processes where two photons from the same excitation pulse or two subsequent excitation pulses were interacting with the QD. $\Delta t$ is the separation between excitation pulses.

\subsubsection{Temporal Correlations}

A standard way of estimating entanglement in a bipartite system $\ket{\psi}_{A, B} = \sum_i \lambda_i \ket{i}_A \ket{i}_B$ is via the purity of the reduced density matrix $\Tr(\rho_A^2) = \sum_i \lambda_i^4$. 
For a maximally entangled state $\Tr(\rho_A^2)=1/N$ (where $N$ is the dimension of $\rho_A$), while for a separable state $\Tr(\rho_A^2)=1$.
While we do not have experimental access to the phase information from $C^{(2)}_{tt}(t_1,t_2)$, we can instead quantify the temporal intensity correlation, which introduces a bound on the purity.

To extract the temporal correlations of the time-resolved coincidence counts $C^{(2)}_{tt}(t_1,t_2)$ in Fig.~\ref{figure2}, we do a Schmidt decomposition of the matrix containing the square root of the count rates $C'_{jl} = \sqrt{C^{(2)}_{tt}(d_t j, d_t l)}$, where $d_t$ the time bin size. 
We perform a singular value decomposition of $C'$, obtaining $C'=\sum_i\lambda_i v_i u_i^\dag$ with $\lambda_i$ the singular values of $C'$ (normalized as $\sum_i\lambda_i^2=1$) and $u_i$, $v_i$ unitary matrices. 
We then use the obtained singular values $\lambda_i$ to estimate the temporal correlation of $C'$ via the quantity $T_c=1-\sum_i\lambda_i^4$ defined in the main text \cite{zielnicki2018joint,Fang2014}. 
This quantifies the degree of temporal correlations in $C^{(2)}_{tt}(t_1,t_2)$ such that $T_c\sim 0$ implies the uncorrelated case (the matrix can be factorized $C'_{jl}=|\Psi^{(1)}_t(d_t j)| |\Psi^{(1)}_t(d_t l)|$) and $T_c\sim 1$ corresponds the maximally correlated case.

In practice, the value of $T_c$ is sensitive to the time bin size $d_t$.
To enable a fair comparison between data sets of different pulse widths, we must therefore vary $d_t$ independently for each data set.
To do this, for each data set $C^{(2)}_{tt}$ we calculate the maximum count value in any bin $c_\text{max}$ and then take the mean across all data sets $\bar{c}_\text{max}$ to give a target count value.
For each data set we then increase $d_t$ until there is at least a single element of $C^{(2)}_{tt}$ with a count value greater than $\bar{c}_\text{max}$.
We repeat this analysis independently for the data of the correlated ($\Delta t\sim 0$) and uncorrelated scattering ($\Delta t\gg \tau$), for which we have $C'_{jl}\approx |\Psi^{(2)}_{tt}(d_t j,d_t l)|+{\cal O}(|\alpha|^2) $ and $C'_{jl}\approx |\Psi^{(1)}_{t}(d_t j)| |\Psi^{(1)}_{t}(d_t l)|+{\cal O}(|\alpha|^2)$, respectively.
Error bars are estimated by performing a Monte Carlo analysis on the entire data processing pipeline, assuming Poissonian distributed count rates.

\pagebreak
\widetext
\begin{center}
\textbf{\large Supplemental Material}
\end{center}
\setcounter{equation}{0}
\setcounter{figure}{0}
\setcounter{table}{0}
\setcounter{page}{1}
\makeatletter
\renewcommand{\theequation}{S\arabic{equation}}
\renewcommand{\thefigure}{S\arabic{figure}}
\renewcommand{\bibnumfmt}[1]{[S#1]}
\renewcommand{\citenumfont}[1]{S#1}

\section{Transmission spectra of the probe photon, under the influence of the control photon}

\begin{figure*}[ht]
	\includegraphics[width=0.9\linewidth]{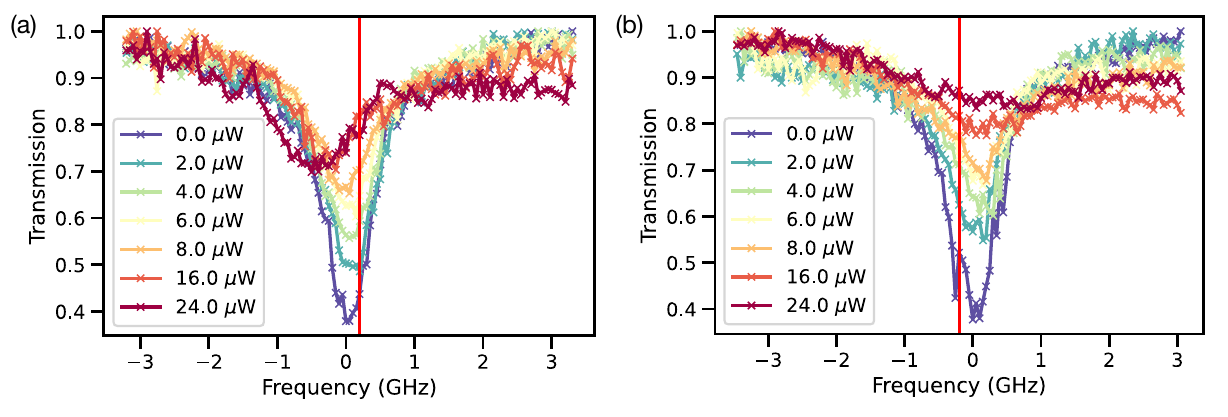}
	\caption{Probe transmission spectra, under the influence of a (a) $\delta_c = 0.3\Gamma$- and (b) $\delta_c = -0.3\Gamma$-detuned control field for different input powers. The full red line indicates the frequency detuning of the control field.
		}
	\label{SM_Fig_Trans_2color}
\end{figure*}

In this section we present in Fig. \ref{SM_Fig_Trans_2color} the evolution of the transmission spectra of the probe field, under the action of the control field detuned by (a) $\delta_c = 0.3\Gamma$ and (b) $\delta_c = -0.3\Gamma$, for different average control powers. By using the scaling presented in Methods, we can extract the frequency shift evolution with the scaled photon flux $n_{\tau}$, as is presented in Fig. 1. (d) in the main text.

\section{Link between the nonlinear two-photon response and coincidence measurements}

To study the two-photon nonlinear response of the QD, we shine a weak coherent wavepacket in the transmission direction which reads \cite{Ramos2017,LeJeannic2020}
\begin{align}
   |\Psi_{\rm in}^{(\alpha)}\rangle = |0\rangle + \alpha |\Psi_{\rm in}^{(1)}\rangle+\frac{\alpha^2}{\sqrt{2}}|\Psi_{\rm in}^{(2)}\rangle+{\cal O}(\alpha^3),\label{weakInputSM}
\end{align}
where $|\alpha|^2\ll 1$ is the weak flux per unit lifetime of the pulse. This pulse is a superposition of a single- $|\Psi_{\rm in}^{(1)}\rangle$ and two-photon $|\Psi_{\rm in}^{(2)}\rangle$ wavepacket components which read
\begin{align}
|\Psi_{\rm in}^{(1)}\rangle ={}& \int d\omega \tilde{f}(\omega)|1_\omega^t\rangle,\label{SingleInput}\\
|\Psi_{\rm in}^{(2)}\rangle={}&\iint \frac{d\omega_1 d\omega_2}{\sqrt{2}} \tilde{f}(\omega_1)\tilde{f}(\omega_2)|1_{\omega_1}^t\rangle |1_{\omega_2}^t\rangle,\label{TwoInput}
\end{align}
where $|1_\omega^t\rangle$ is a monochromatic single-photon Fock state of frequency $\omega$ propagating in forward ($\mu=t$) direction and $\bar{f}(\omega)=(\delta t^2/\pi)^{(1/4)}e^{-\delta t^2(\omega-\omega_0)^2/2}\sum_{n=1}^{N_p}e^{i(\omega-\omega_0)T(n-1)}$ is the frequency profile of the train of $N_p$ Gaussian wavepackets centered at the QD transition frequency $\omega_0$, with time width $\delta t$, and temporal separation $T$ between pulses. The Fourier transform $f(t)=(2\pi)^{-1/2}\int d\omega e^{-i\omega t}\tilde{f}(\omega)=(\pi\delta t^2)^{-1/4}\sum_{n=1}^{N_p}e^{-(t-[n-1]T)^2}$ is the temporal profile of the train of Gaussian wavepackets.

This train of coherent Gaussian wavepackets interact with the QD and we measure coincidence counts $C_{\mu\mu'}^{(2)}(t_1,t_2)$ at each output port of the waveguide $\mu,\mu'=t,r$ with two detectors clicking at times $t_1$ and $t_2$. These measurements are proportional to the unnormalized second-order correlation function defined as
\begin{align}
G^{(2)}_{\mu\mu'}(\tau) = \frac{\langle \Psi_{\rm in}^{(\alpha)}| a_{\rm out}^{\mu}{}^\dag(t) a_{\rm out}^{\mu'}{}^\dag(t+\tau) a_{\rm out}^{\mu'}(t+\tau) a_{\rm out}^{\mu}(t) |\Psi_{\rm in}^{(\alpha)}\rangle_{\rm ss}}{|\alpha|^4},\label{G2def}
\end{align}
where $a_{\rm out}^{\mu}(t)=S a_{\rm in}^{\mu}(t) S^\dag$ is the annihilation operator of a photon detected at time $t$ at output channel $\mu$ after interacting with the QD, and $S$ is the Scattering matrix of the quantum dot. Replacing Eqs.~(\ref{weakInputSM})-(\ref{TwoInput}) into Eq.~(\ref{G2def}), we find that up to small corrections on the flux per lifetime $|\alpha|^2\ll 1$, the two-time correletion function $G^{(2)}_{\mu\mu'}(t_1,t_2)$ can be interpreted as a two-photon wavefunction,
\begin{align}
G^{(2)}_{\mu\mu'}(\tau) = |\Psi^{(2)}_{\mu\mu'}(t_1,t_2)|^2 + {\cal O}(|\alpha|^2),
\end{align}
with $\Psi^{(2)}_{\mu\mu'}(t_1,t_2)= (2\pi)^{-1}\iint d\omega_1 d\omega_2 e^{-i(\omega_1 t_1+\omega_2 t_2)}\langle 1_{\omega_1}^{\mu}, 1_{\omega_2}^{\mu'}|\Psi^{(2)}_{\rm out}\rangle/\sqrt{2}$. This two-photon wavefunction can be further decomposed as,
\begin{align}
    \Psi^{(2)}_{\mu\mu'}(t_1,t_2) = \Psi^{(1)}_{\mu}(t_1)\Psi^{(1)}_{\mu'}(t_2)+N(t_1,t_2),
\end{align}
where $\Psi^{(1)}_{\mu}(t)=(2\pi)^{-1/2}\int d\omega e^{-i\omega t}\langle 1^\mu_\omega|\Psi^{(1)}_{\rm out}\rangle$ at direction $\mu=t,r$. In addition, the two-photon nonlinearity $N(t_1,t_2)$ is given by a convolution of the input wavepacket profile $\tilde{f}(\omega)$ and the nonlinear part of the scattering matrix ${\cal T}_{\omega_1\omega_2}(t_2-t_1)$ \cite{LeJeannic2020} as
\begin{align}
    N(t_1,t_2) ={}& \frac{1}{2\pi}\int d\omega_1 d\omega_2 \tilde{f}(\omega_1)\tilde{f}(\omega_2)e^{-i(\omega_1+\omega_2)(t_1+t_2)/2}{\cal T}_{\omega_1\omega_2}(t_2-t_1).
\end{align}
In the case of a single two-level emitter with lifetime $\tau$, total decay rate $\gamma=1/\tau$, dephasing rate $\Gamma_0$, decay into the waveguide modes $\gamma_{\rm wg}$, and coupling efficiency into the waveguide $\beta=\gamma_{1D}/\gamma$, the explicit expression of the two-photon nonlinearity reads \cite{LeJeannic2020}, 
\begin{align}
    N(t_1,t_2)={}&-\frac{\beta^2}{2\pi}e^{-(\gamma/2+\Gamma_0)|t_2-t_1|}\left( \int d\omega e^{-i(\omega/2)(t_1+t_2-|t_1-t_2|)}\tilde{f}(\omega)\frac{\gamma/2}{\gamma/2+\Gamma_0-i(\omega-\omega_0)}\right)^2.\label{explicitN}
\end{align}

Evaluating Eq.~(\ref{explicitN}) we see that for $|t_2-t_1|\gg \tau$, the nonlinearity vanishes $N(t_1,t_2)\sim 0$ and the two-photon wavefunction factorizes $C^{(2)}_{\mu\mu'}(t_1,t_2)\propto |\Psi^{(2)}_{\mu\mu'}(t_1,t_2)|^2\sim |\Psi^{(1)}_{\mu}(t_1)|^2|\Psi^{(1)}_{\mu'}(t_2)^2|$, as we observe in the data of the main text. For $|t_2-t_1|\lesssim \tau$ we have an appreciable two-photon nonlinearity $N(t_1,t_2)$ and thus complex two-photon correlations are described by $C^{(2)}_{\mu\mu'}(t_1,t_2)\propto |\Psi^{(2)}_{\mu\mu'}(t_1,t_2)|^2= |\Psi^{(1)}_{\mu}(t_1)\Psi^{(1)}_{\mu'}(t_2)+N(t_1,t_2)|^2$.

\section{Complete data set}
\begin{figure*}[ht]
	\includegraphics[width=0.9\linewidth]{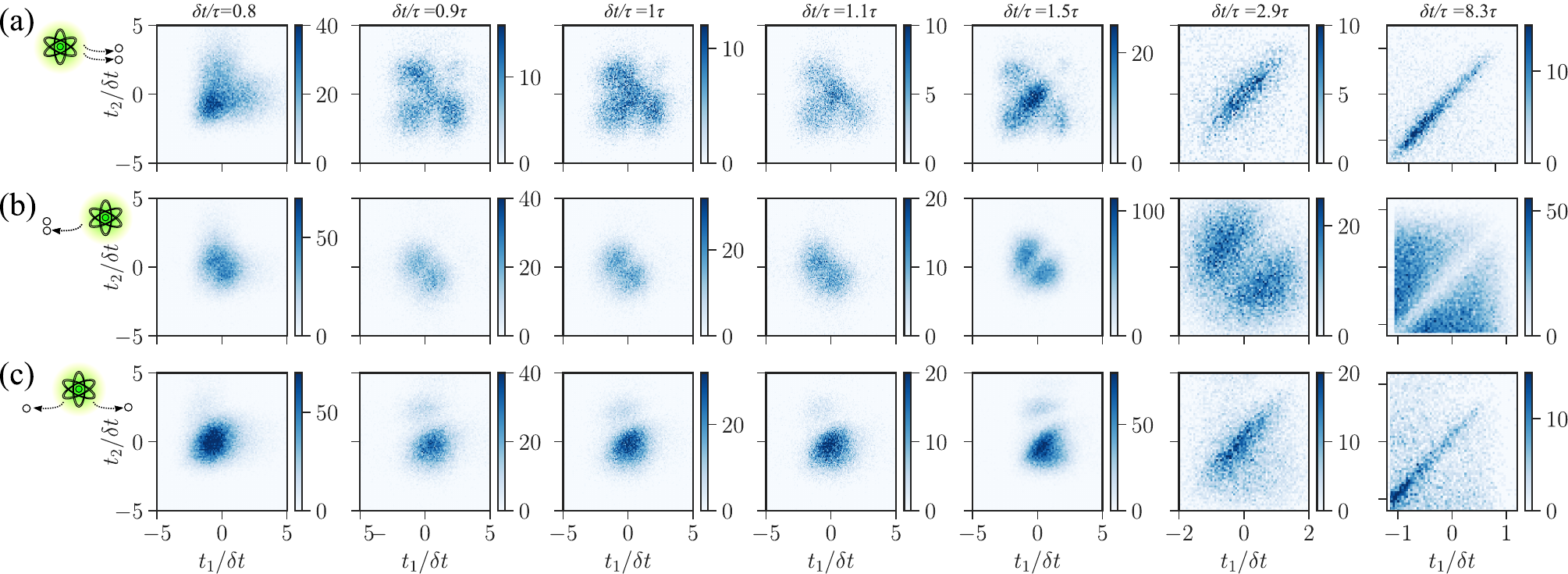}
	\caption{Measured time-resolved second order correlation data for transmitted (a), reflected (b) and transmitted reflected pairs of photons,belonging to the same excitation wavepacket.
		}
	\label{SM_Fig_LpB}
\end{figure*}

\begin{figure*}[ht]
	\includegraphics[width=0.9\linewidth]{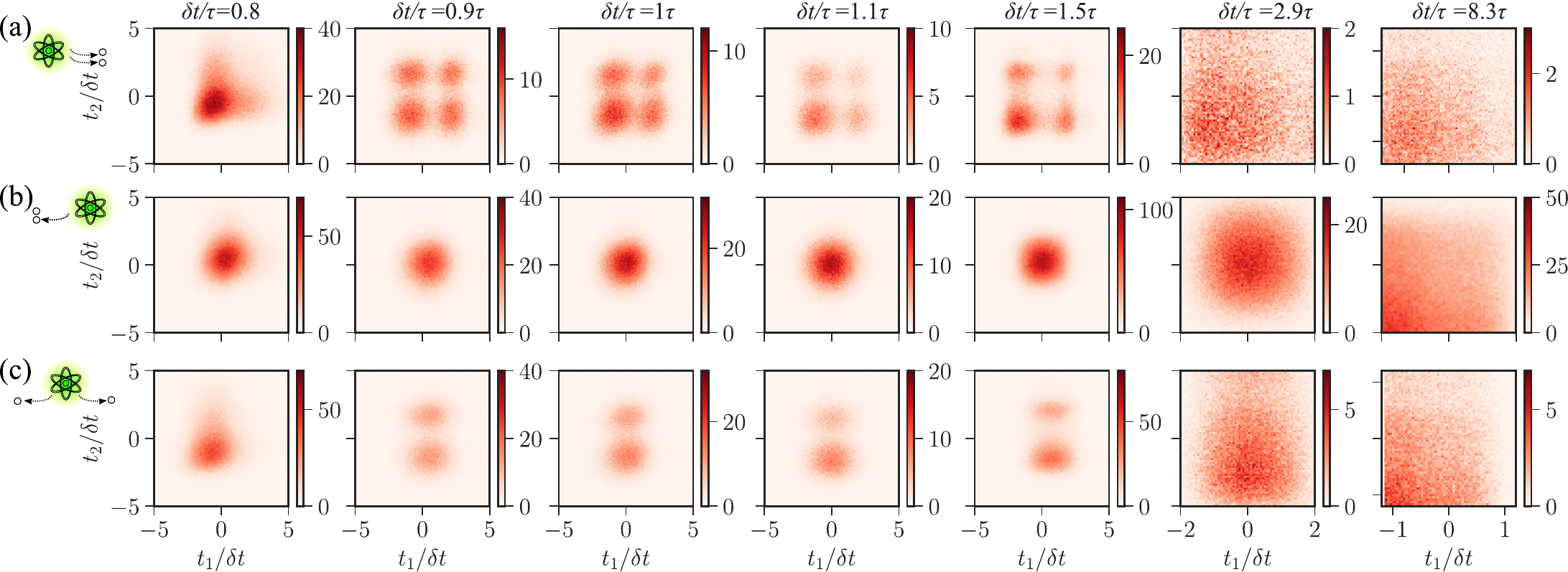}
	\caption{Measured time-resolved second order correlation data for transmitted (a), reflected (b) and transmitted reflected pairs of photons, belonging to two different excitation wavepackets.
		}
	\label{SM_Fig_L}
\end{figure*}

In this section we present complete dataset, containing the measured time-resolved second order correlation data for the correlated (Fig. \ref{SM_Fig_LpB}) and uncorrelated (Fig. \ref{SM_Fig_L}) two-photon wavepacket input, for both output photons transmitted (a), reflected (b) or going in opposite direction (c).

\begin{figure*}[ht]
	\includegraphics[width=0.9\linewidth]{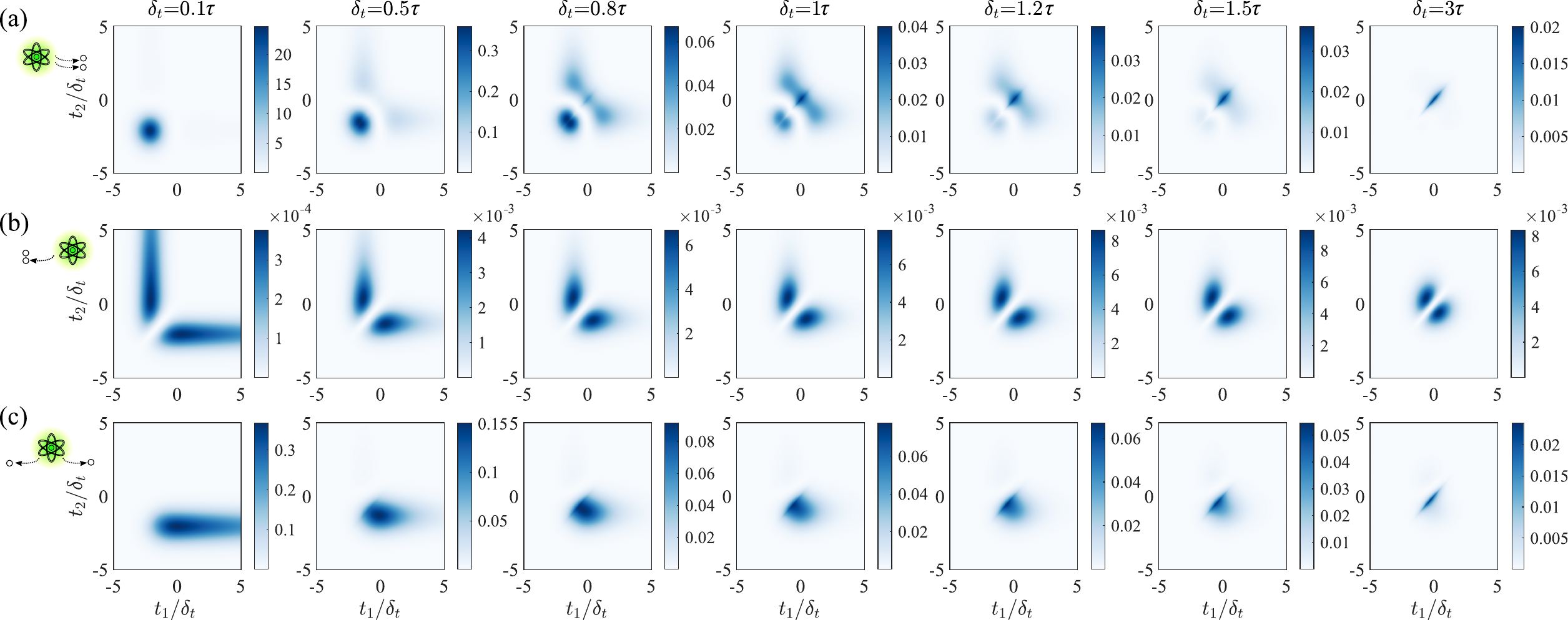}
	\caption{Calculated time-resolved second order correlation data for transmitted (a), reflected (b) and transmitted reflected (c) pairs of photons, belonging to the same excitation wavepacket.
		}
	\label{SM_Fig_thLpB}
\end{figure*}

\begin{figure*}[ht]
	\includegraphics[width=0.9\linewidth]{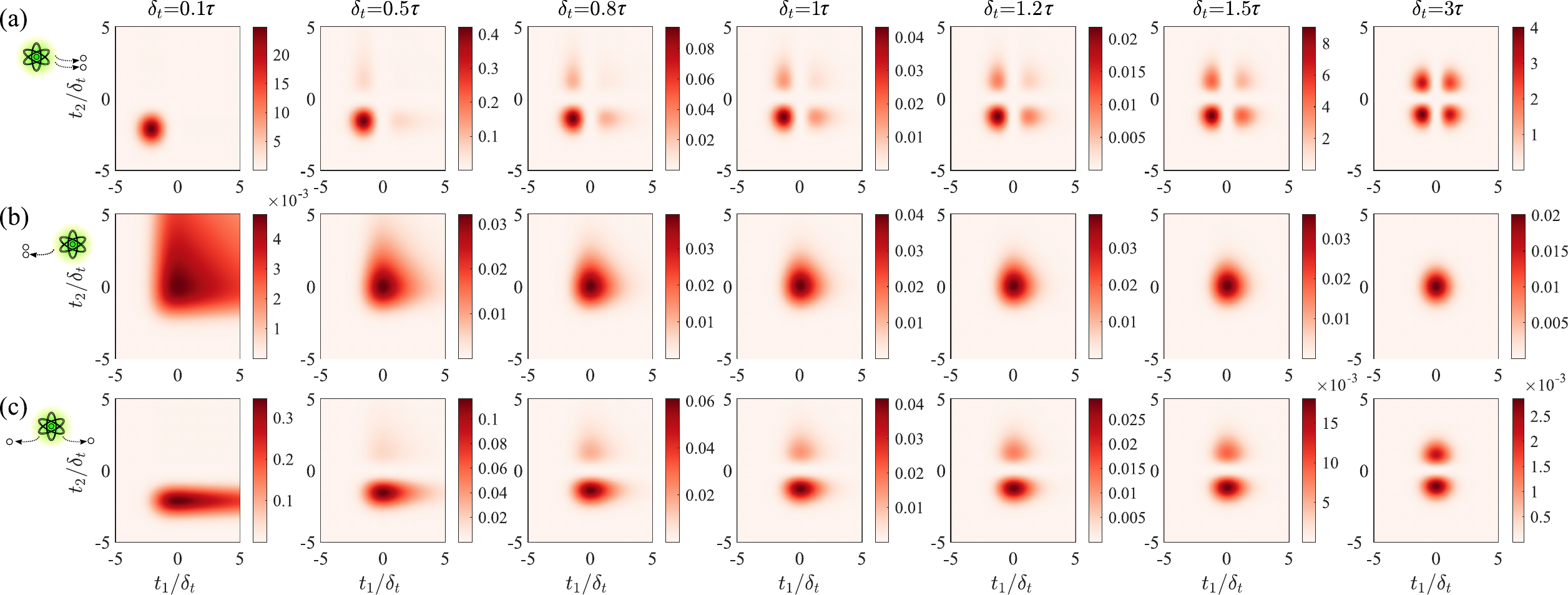}
	\caption{Calculated time-resolved second order correlation data for transmitted (a), reflected (b) and transmitted reflected (c) pairs of photons, belonging to two different excitation wavepackets.
		}
	\label{SM_Fig_thL}
\end{figure*}

We also present, in Fig.~ \ref{SM_Fig_thLpB} and Fig.~ \ref{SM_Fig_thL}, corresponding calculated correlations, in the ideal case of $\beta=1$ and in the absence of dephasing. We have also extended the range of pulse width to much shorter pulses for a better understanding of the dynamics.

\end{document}